\documentstyle[12pt,qqa4lart]{article}
\input tcilatex

\QQQ{Language}{
American English
}

\begin{document}

\author{Lu-Ming Duan and Guang-Can Guo\thanks{%
Electronic address: gcguo@sunlx06.nsc.ustc.edu.cn} \\
Department of Physics and Nonlinear Science Center,\\
University of Science and Technology of China,\\
Hefei 230026, People's Republic of China}
\title{Ghost interference and diffraction based on the beam splitter}
\date{}
\maketitle

\begin{abstract}
\baselineskip 20ptA simple scheme is proposed for observing the ghost
interference and diffraction. The signal and the idler beams are produced by
a beam splitter with the incident light being in a thermal state. A slit is
inserted into the {\it signal} beam. We derive rigorously that
interference-diffraction patterns can be observed in the first-order
correlation by scanning the probe in the {\it idler} beam.\\

\end{abstract}

\newpage\ \baselineskip 20ptQuantum correlation was recognized as one of the
most striking features of quantum mechanics ever since Einstein, Podolsky,
and Rosen proposed their famous {\it gedanken experiment }[1]. In various
applications of non-classical features, such as in quantum cryptography
[2-4] and in teleportation [5,6], correlations of states play an essential
role . In recent years, quantum correlation has been confirmed in a number
of two-photon correlation experiments [7-15]. Among those, the ghost
interference-diffraction [15] is a remarkable example. In the original
observation [15] of ghost diffraction, a correlated two-photon state is
generated in the beta barium borate (BBO) crystal by the process of
spontaneous parametric down-conversion (SPDC). The SPDC light beam, which
consists of two orthogonal polarization components (usually called signal
and idler), is split by a polarization beam splitter into two beams, and
detected by two distant pointlike photon counting detectors for
coincidences. A Young's double-slit or single-slit aperture is inserted into
the {\it signal} beam. Surprisingly, an interference-diffraction pattern is
observed in the coincidence counts by scanning the detector in the {\it idler%
} beam. This is called the ghost interference-diffraction, whose most
striking feature is that the slit is inserted into the signal beam and in
the idler beam ''appears'' the interference-diffraction pattern.

The ghost interference-diffraction results from quantum correlation. In the
original experiment [15], the correlation is produced by the SPDC process.
Beam splitters can also be used to generate quantum correlation [16]. In
this letter, we propose a simple scheme for observing the ghost interference
and diffraction. The scheme is based on the correlation generated by a beam
splitter. We use the thermal light source and measure the first-order
correlation between the signal and the idler beams. This is contrast to the
original experiment, where one in fact measures the second-order
correlation. (Note that for a correlated two-photon state the coincidence
counting rate is proportional to the second-order correlation function
between the signal and the idler beams.) The probable experimental setup for
this scheme is illustrated in Fig. 1. \\

\begin{center}
Fig. 1\\
\end{center}

In Fig. 1, $L_1,$ $L_2$ and $L_3$ are convex lenses. The line thermal light
source $S$ is put at the focal plane of lens $L_1$, so after $L_1$ the beam
has considerably large angular uncertainty. The beam is split by a beam
splitter (BS) into two correlated beams, the signal beam and the idler beam.
A single- or double-slit is inserted into the signal beam. After lens $L_2$,
the signal light is coupled to a fiber. The input tip of the fiber is fixed
on the axis at the focal plane of $L_2$. The idler light, passing through
lens $L_3$, is coupled to another fiber. The horizontal transverse
coordinate $x$ of the fiber input tip is scanned by an encoder driver at the
focal plane of $L_3$. After a delay $\tau $, the idler light and the signal
light are superposed and then detected by a photon-counting detector $D$.
The visibility of the interference fringe gives the first-order correlation
between the signal and the idler beams. By scanning the fiber input tip in
the idler beam, an interference-diffraction pattern of the slit inserted
into the signal beam will occur in the first-order correlation.

In the following, we derive the interference-diffraction pattern of the slit
in the first-order correlation. In this scheme, the input light of the beam
splitter is a multi-mode thermal optical field. Different modes of the
thermal light are independent of each other. The mode with angular $%
\overrightarrow{k}=\left( k_x,k_y,\sqrt{\frac{\omega ^2}{c^2}-k_x^2-k_y^2}%
\right) $ is denoted by $a_{\overrightarrow{k}}$, whose corresponding output
modes, the signal mode and the idler mode, are denoted by $b_{%
\overrightarrow{k}}$ and $c_{\overrightarrow{k}}$, respectively. Another
corresponding input mode of the beam splitter, which is in the vacuum state,
is denoted by $a_{\overrightarrow{k}}^{^{\prime }}$. The input-output
theory, applied to the beam splitter, yields the following canonical
transformation [17] 
\begin{equation}
\label{1}\left( 
\begin{array}{c}
b_{
\overrightarrow{k}} \\ c_{\overrightarrow{k}}
\end{array}
\right) =\left( 
\begin{array}{cc}
r & t \\ 
-t & r
\end{array}
\right) \left( 
\begin{array}{c}
a_{
\overrightarrow{k}} \\ a_{\overrightarrow{k}}^{^{\prime }}
\end{array}
\right) ,
\end{equation}
where the parameters $r$ and $t$ satisfy $r^2+t^2=1$. For a $50-50\%$ beam
splitter, $r=t=\frac 1{\sqrt{2}}$. From Eq. (1), we easily obtain the
relation between the normal characteristic functions for the output modes
and for the input modes 
\begin{equation}
\label{2}
\begin{array}{c}
\chi ^{\left( n\right) }\left( b_{
\overrightarrow{k}},c_{\overrightarrow{k}};\xi _{1\overrightarrow{k}},\xi _{2%
\overrightarrow{k}}\right) =\left\langle \exp \left[ i\left( \xi _{1%
\overrightarrow{k}}^{*}b_{\overrightarrow{k}}^{\dagger }+\xi _{2%
\overrightarrow{k}}^{*}c_{\overrightarrow{k}}^{\dagger }\right) \right] \exp
\left[ i\left( \xi _{1\overrightarrow{k}}b_{\overrightarrow{k}}+\xi _{2%
\overrightarrow{k}}c_{\overrightarrow{k}}\right) \right] \right\rangle  \\  
\\ 
=\chi ^{\left( n\right) }\left( a_{\overrightarrow{k}},a_{\overrightarrow{k}%
}^{^{\prime }};r\xi _{1\overrightarrow{k}}-t\xi _{2\overrightarrow{k}},t\xi
_{1\overrightarrow{k}}+r\xi _{2\overrightarrow{k}}\right) .
\end{array}
\end{equation}
We use the symbol $\chi ^{\left( n\right) }\left( \left\{ a_{\overrightarrow{%
k}}\right\} ;\left\{ \xi _{\overrightarrow{k}}\right\} \right) $ to denote
the total normal characteristic function of all the modes $a_{%
\overrightarrow{k}}$ with variables $\xi _{\overrightarrow{k}}$,
respectively. Since all the modes $a_{\overrightarrow{k}}$ are in thermal
states and independent of each other, the characteristic function $\chi
^{\left( n\right) }\left( \left\{ a_{\overrightarrow{k}}\right\} ;\left\{
\xi _{\overrightarrow{k}}\right\} \right) $ has the form [18] 
\begin{equation}
\label{3}\chi ^{\left( n\right) }\left( \left\{ a_{\overrightarrow{k}%
}\right\} ;\left\{ \xi _{\overrightarrow{k}}\right\} \right) =\stackunder{%
\overrightarrow{k}}{\prod }\exp \left( -\left| \xi _{\overrightarrow{k}%
}\right| ^2\left\langle N_{\overrightarrow{k}}\right\rangle \right) ,
\end{equation}
where $\left\langle N_{\overrightarrow{k}}\right\rangle $ represents the
mean photon number of the mode $a_{\overrightarrow{k}}$. The function $%
\left\langle N_{\overrightarrow{k}}\right\rangle $ with variable $%
\overrightarrow{k}$ determines the intensity distribution of the beam among
different directions. Note that all the modes $a_{\overrightarrow{k}%
}^{^{\prime }}$ are in the vacuum state. Combining Eqs. (2) and (3), we thus
get the following total normal characteristic function for all the modes $b_{%
\overrightarrow{k}}$ and $c_{\overrightarrow{k}}$%
\begin{equation}
\label{4}\chi ^{\left( n\right) }\left( \left\{ b_{\overrightarrow{k}%
}\right\} ,\left\{ c_{\overrightarrow{k}}\right\} ;\left\{ \xi _{1%
\overrightarrow{k}}\right\} ,\left\{ \xi _{2\overrightarrow{k}}\right\}
\right) =\exp \left\{ -\stackunder{\overrightarrow{k}}{\sum }\left[
\left\langle N_{\overrightarrow{k}}\right\rangle \left| r\xi _{1%
\overrightarrow{k}}-t\xi _{2\overrightarrow{k}}\right| ^2\right] \right\} .
\end{equation}

The signal beam passes through a single- or double-slit aperture. This is a
Fraunhofer diffraction. The diffraction mode with angular $\overrightarrow{k}%
^{^{\prime }}=\left( k_x^{^{\prime }},k_y^{^{\prime }},\sqrt{\frac{\omega ^2%
}{c^2}-k_x^{^{\prime }2}-k_y^{^{\prime }2}}\right) $ is denoted by $d_{%
\overrightarrow{k}^{^{\prime }}}$. To determine the state of the diffraction
modes, we need a quantum formalism of Fraunhofer diffraction. In the early
days of quantum electrodynamics, it had been proven that the Maxwell
equations which underpin diffraction remain true when the fields are
quantized [19]. In quantum optics, the entire mode structure of the
diffraction field is still determined by the Helmoholtz part of the wave
equation. The role played by quantum mechanics is in determining the state
of the diffraction modes from that of the incident modes. In a recent paper
[20], by introducing quantum correspondence of the Kirchhoff boundary
condition, we developed a simple method to determine the state of the
diffraction modes. The formalism is based on the normal characteristic
functions. The quantum Kirchhoff boundary condition states as follows: When
an optical field passes through a diffraction plane, the modes at the
aperture undergo no dissipation, whereas the modes at the plane undergo such
a strong dissipation that after the screen they are all in the vacuum state.
Starting from this boundary condition, we rigorously derive that the normal
characteristic function of the diffraction modes can be connected with that
of the incident modes by the following equation 
\begin{equation}
\label{5}\chi ^{\left( n\right) }\left( \left\{ d_{\overrightarrow{k}%
^{^{\prime }}}\right\} ;\left\{ \xi _{\overrightarrow{k}^{^{\prime
}}}\right\} \right) =\chi ^{\left( n\right) }\left( \left\{ b_{%
\overrightarrow{k}}\right\} ;\left\{ \sqrt{\lambda }\stackunder{%
\overrightarrow{k}^{^{\prime }}}{\sum }\left[ \xi _{\overrightarrow{k}%
^{^{\prime }}}f\left( \overrightarrow{k}^{^{\prime }}-\overrightarrow{k}%
\right) \right] \right\} \right) ,
\end{equation}
where the function $f\left( \overrightarrow{k}\right) $ is the Fraunhofer
diffraction factor, defined as 
\begin{equation}
\label{6}f\left( \overrightarrow{k}\right) =\frac{\sqrt{\lambda }}\Sigma
\int_\Sigma e^{-i\left( k_xx+k_yy\right) }dxdy.
\end{equation}
and normalized by 
\begin{equation}
\label{7}\stackunder{\overrightarrow{k}}{\sum }\left| f\left( 
\overrightarrow{k}\right) \right| ^2=1,
\end{equation}
in which the symbol $\stackunder{\overrightarrow{k}}{\sum }$ stands for $%
\stackunder{k_x,k_y}{\sum }$. The parameters $\lambda $ in Eqs. (5) and (6)
is the energy transmissivity, defined by $\lambda =\frac \Sigma S$, whose
physical meaning is the ratio of total energy of the diffraction field to
that of the incident field. $\Sigma $ and $S$ denote areas of the
diffraction aperture and of the whole diffraction plane, respectively. Since 
$\Sigma $ may represent a single- or double or $N$-slit, Eq. (5) gives
quantum formalism for interference as well as for diffraction. The
Fraunhofer diffraction factor determines the interference-diffraction
pattern. For example, for an $N$-slit aperture (in $X$-axis direction), $%
f\left( \overrightarrow{k}\right) $ reads explicitly 
\begin{equation}
\label{8}f\left( \overrightarrow{k}\right) =\sqrt{\lambda }\frac{\sin \left(
k_xa/2\right) }{k_xa/2}\frac{\sin \left( Nk_xd/2\right) }{k_xd/2}\delta
\left( k_y\right) ,
\end{equation}
where $a$ and $d$ are the slit width and slit distance, respectively.

Eqs. (4) and (5), combined together, yield the normal characteristic
function for all the modes $d_{\overrightarrow{k}^{^{\prime }}}$ and $c_{%
\overrightarrow{k}}$%
\begin{equation}
\label{9}
\begin{array}{c}
\chi ^{\left( n\right) }\left( \left\{ d_{
\overrightarrow{k}^{^{\prime }}}\right\} ,\left\{ c_{\overrightarrow{k}%
}\right\} ;\left\{ \xi _{\overrightarrow{k}^{^{\prime }}}\right\} ,\left\{
\xi _{2\overrightarrow{k}}\right\} \right)  \\  \\ 
=\chi ^{\left( n\right) }\left( \left\{ b_{
\overrightarrow{k}}\right\} ,\left\{ c_{\overrightarrow{k}}\right\} ;\left\{ 
\sqrt{\lambda }\stackunder{\overrightarrow{k}^{^{\prime }}}{\sum }\left[ \xi
_{\overrightarrow{k}^{^{\prime }}}f\left( \overrightarrow{k}^{^{\prime }}-%
\overrightarrow{k}\right) \right] \right\} ,\left\{ \xi _{2\overrightarrow{k}%
}\right\} \right)  \\  \\ 
=\exp \left\{ -\stackunder{\overrightarrow{k}}{\sum }\left[ \left\langle N_{%
\overrightarrow{k}}\right\rangle \left| r\sqrt{\lambda }\stackunder{%
\overrightarrow{k}^{^{\prime }}}{\sum }\left[ \xi _{\overrightarrow{k}%
^{^{\prime }}}f\left( \overrightarrow{k}^{^{\prime }}-\overrightarrow{k}%
\right) \right] -t\xi _{2\overrightarrow{k}}\right| ^2\right] \right\} .
\end{array}
\end{equation}
We measure the first-order correlation between a diffraction mode $d_{%
\overrightarrow{k}_0^{^{\prime }}}$ (fixed) and an idler mode $c_{%
\overrightarrow{k}}$ (varying). From Eq. (9), the correlation function
between $d_{\overrightarrow{k}_0^{^{\prime }}}$ and $c_{\overrightarrow{k}}$
is 
\begin{equation}
\label{10}
\begin{array}{c}
\left\langle c_{
\overrightarrow{k}}^{\dagger }d_{\overrightarrow{k}_0^{^{\prime
}}}\right\rangle =\left. \frac{\partial ^2}{\partial \xi _{2\overrightarrow{k%
}}^{*}\partial \xi _{\overrightarrow{k}_0^{^{\prime }}}}\chi ^{\left(
n\right) }\left( \left\{ d_{\overrightarrow{k}^{^{\prime }}}\right\}
,\left\{ c_{\overrightarrow{k}}\right\} ;\left\{ \xi _{\overrightarrow{k}%
^{^{\prime }}}\right\} ,\left\{ \xi _{2\overrightarrow{k}}\right\} \right)
\right| _{\xi _{2\overrightarrow{k}}=\xi _{\overrightarrow{k}_0^{^{\prime
}}}=0} \\  \\ 
=rt\sqrt{\lambda }\left\langle N_{\overrightarrow{k}}\right\rangle f\left( 
\overrightarrow{k}_0^{^{\prime }}-\overrightarrow{k}\right) .
\end{array}
\end{equation}
Hence the first-order degree of correlation $g^{\left( 1\right) }$ reads 
\begin{equation}
\label{11}g^{\left( 1\right) }\left( \overrightarrow{k}\right) =\frac{%
\left\langle c_{\overrightarrow{k}}^{\dagger }d_{\overrightarrow{k}%
_0^{^{\prime }}}\right\rangle }{\sqrt{\left\langle c_{\overrightarrow{k}%
}^{\dagger }c_{\overrightarrow{k}}\right\rangle \left\langle d_{%
\overrightarrow{k}_0^{^{\prime }}}^{\dagger }d_{\overrightarrow{k}%
_0^{^{\prime }}}\right\rangle }}=\frac{\sqrt{\left\langle N_{\overrightarrow{%
k}}\right\rangle }f\left( \overrightarrow{k}_0^{^{\prime }}-\overrightarrow{k%
}\right) }{\sqrt{\stackunder{\overrightarrow{k}^{^{\prime \prime }}}{\sum }%
\left[ \left\langle N_{\overrightarrow{k}^{^{\prime \prime }}}\right\rangle
\left| f\left( \overrightarrow{k}_0^{^{\prime }}-\overrightarrow{k}%
^{^{\prime \prime }}\right) \right| ^2\right] }}.
\end{equation}
If the factor $f\left( \overrightarrow{k}\right) $ varies much faster than
the distribution $\left\langle N_{\overrightarrow{k}}\right\rangle $, from
Eq. (7), $g^{\left( 1\right) }\left( \overrightarrow{k}\right) $
approximately equals $f\left( \overrightarrow{k}_0^{^{\prime }}-%
\overrightarrow{k}\right) $. This is an interesting phenomenon. The
correlation $g^{\left( 1\right) }\left( \overrightarrow{k}\right) $ can be
obtained by measuring the visibility of the interference pattern between the
fixed diffraction mode $d_{\overrightarrow{k}_0^{^{\prime }}}$ and a varying
idler mode $c_{\overrightarrow{k}}$. Eq. (11) shows the measurement result
reveals the interference-diffraction pattern of the slit. In the setup
illustrated by Fig. 1, the fiber input tip is fixed on the axis in the
signal beam and scanned in $X$-axis direction in the idler beam, so $%
k_{0x}^{^{\prime }}=k_{0y}^{^{\prime }}=k_y=0$ and $k_x\approx \frac{2\pi x}{%
\lambda f_3}$, where $\lambda =\frac{2\pi }{\left| \overrightarrow{k}\right| 
}=\frac{2\pi c}\omega $ and $f_3$ is the focal distance of lens $L_3$.
Substituting these expressions into Eqs. (11) and (8), we get 
\begin{equation}
\label{12}g^{\left( 1\right) }\left( x\right) \approx \sqrt{\lambda }\frac{%
\sin \left( \frac{\pi xa}{\lambda f_3}\right) }{\frac{\pi xa}{\lambda f_3}}%
\frac{\sin \left( \frac{\pi Nxd}{\lambda f_3}\right) }{\frac{\pi xd}{\lambda
f_3}}.
\end{equation}
Thus the interference-diffraction pattern of the slit inserted into the
signal beam occurs in the first-order correlation by scanning the fiber
input tip in the idler beam. This is the ghost interference-diffraction.

A notable fact in this experimental scheme is that if we measure intensity
of the signal beam rightly after the slit, no interference-diffraction
patterns appear. This follows from the equation 
\begin{equation}
\label{13}\left\langle d_{\overrightarrow{k}^{^{\prime }}}^{\dagger }d_{%
\overrightarrow{k}^{^{\prime }}}\right\rangle =r^2\lambda \stackunder{%
\overrightarrow{k}}{\sum }\left[ \left\langle N_{\overrightarrow{k}%
}\right\rangle \left| f\left( \overrightarrow{k}^{^{\prime }}-%
\overrightarrow{k}\right) \right| ^2\right] \approx r^2\lambda \left\langle
N_{\overrightarrow{k}^{^{\prime }}}\right\rangle .
\end{equation}
The absence of the interference-diffraction structure is due to the
considerably large angular propagation uncertainty of the signal beam.

Compared with the original observation of the ghost diffraction, this scheme
has two remarkable features. First, the scheme only involves a common
thermal light source and the correlations are generated simply by a beam
splitter. Second, we measure the first-order correlation between the
diffraction and the idler beams. This can be easily fulfilled by observing
the visibility of the interference pattern between a fixed diffraction mode
and a varying idler mode.\\

{\bf Acknowledgment}

This Project was supported by the National Natural Science Foundation of
China.\newpage\

\end{document}